\documentclass[aps,prl,twocolumn,nofootinbib,groupedaddress,amsfonts,floatfix]{revtex4}

\usepackage{graphicx,amsmath,amssymb,amstext}
\usepackage{amssymb,amsbsy,amsfonts,amsthm}

\begin{document}

\title{Long-lived Positronium and AGN Jets}
\author{John T. Giblin, Jr${}^{1,2,3}$}
\author{Janine Shertzer${}^{4,5}$}
\affiliation{${}^1$Department of Physics, Kenyon College, Gambier, OH  43022}
\affiliation{${}^2$The Perimeter Institute for Theoretical Physics, 31 Caroline St N, Waterloo, ON  N2L 2Y5}
\affiliation{${}^3$Department of Physics, Case Western Reserve University, Cleveland, OH}
\affiliation{${}^4$Department of Physics, College of the Holy Cross, Worcester, MA  01610}
\affiliation{${}^5$ITAMP, Harvard-Smithsonian Center for Astrophysics, 60 Garden Street, Cambridge, MA 02138}

\begin{abstract}
We suggest that stable states of positronium might exist in the jets of active galactic nuclei (AGN).  Electrons and positrons are created near the accretion disks of supermassive black holes at the centers of AGN and are accelerated along magnetic field lines while within the {\sl Alfv\'en radius}.  The conditions in this region are ideal for the creation of bound states of positronium which are stable against annihilation.  Traveling at relativistic speeds along the jet, the helical magnetic field  enables the atoms to survive for great distances.  
\end{abstract}

\maketitle

\section{Introduction}
The current paradigm \cite{1995PASP..107..803U} describes active galactic nuclei (AGN) as the centers of galaxies containing a supermassive, actively accreting, black-hole. In this scenario, the electromagnetic emission from the accretion disk dominates that of the host galaxy.  One particularly interesting characteristic of some AGN is the existence particle jets originating near the accretion disk and directed along the axis of the disk.   Although there is no unique explanation for these jets, evidence suggests \cite{2000AIPC..515...31B} that these jets are leptonic.  

The density of leptons is largest near the black hole, where high energy processes produce a `pair plasma' \cite{1989MNRAS.237..411S,1997A&A...320...26D}.  Once the plasma is ejected along the jet at relativistic speeds \cite{Scarpa2002405}, the density of the plasma decreases with distance.  In this outer region, the density is sufficiently low that electrons and positrons are extremely dilute and annihilation is unlikely.  The critical question is how the electrons and positrons in the high density region near the Aflv\`en radius avoid annihilation.  It has been suggested that annihilation is suppressed because of the relativistic speeds; alternatively, the electrons and positrons annihilate but the high energy gamma rays produce new electron-positron pairs further along in the jet (see for example \cite{RevModPhys.56.255}.) 

The idea that positironium exists in jets is not new \cite{1993MNRAS.263L...9G,PhysRevLett.41.135,1983AIPC..101..314L,1983ApJ...271..804E} and previous authors \cite{1984ApJ...282..291M} have remarked on the difficulty in detecting the characteristic annihilation signal.  Nevertheless, it has been assumed, up until now, that the positronium annihilates quickly.  We propose a novel mechanism by which stable bound states of positronium that are formed near the Alfv\`en radius can travel along the jet and annihilate some distance from the host galaxy. We argue that the astrophysical conditions are ideal for creating and sustaining these states.

The canonical model of jet production breaks the phenomenon into three distinct regions \cite{2010LNP...794..233S}. The first of these is the supermassive black hole accretion disk.  In this region magnetic field lines are anchored on the disk, and forced to rotate along with the matter in the disk.  The anchoring of these lines to the disk causes the magnetic field to be helical on large-scales, even far away from the host galaxy.  

The second region, lying directly above and below the accretion disk is characterized by strong electromagnetic fields.  This region is {\sl force free},
\begin{equation}
\label{forcefree}
\rho \vec{E} + \vec{j}\times \vec{B} = 0.
\end{equation}
meaning that charged particles do not feel the effect of other charged particles.  Such conditions allow for the free production of electron-positron pairs.  

Charged particles can only move freely along the direction of the magnetic field lines.  Since these lines are spinning,  a centrifugal force acts on the electrons and positrons and mechanically accelerates them along the magnetic field lines.  Due to the strength of the external fields, movement perpendicular to the magnetic field lines is circular and not translational.  This creates the jets of particles parallel to the axis of the accretion disk.  This acceleration is quick and both electrons and positrons acquire large velocities before leaving the force-free zone.  Annihilation is unlikely in this region due to the fact that the Coulomb interaction is weak compared to the accelerating mechanism.

Once particles travel far enough away from the accretion disk, the decrease in field strength leads to a breakdown of the force-free condition, Eq.~(\ref{forcefree}).  This occurs, approximately, at the {\sl Alfv\'en radius}  \cite{2010LNP...794..233S}, where the speed of flow approaches the Alfv\'en speed (see \cite{2010LNP...794..233S} for more details).  Phenomenologically, though, we are interested in  where the force-free transition occurs.  Past this surface, the magnetic field becomes almost completely azimuthal; positrons and electrons can now feel the effect of the Coulomb  interaction and annihilation is likely because the density is high.   

We propose that the conditions near the Alfven radius are optimal for creating
stable bound states of positronium.  It is here that equal numbers of positrons and electrons
are traveling at relativistic speeds in a strong magnetic field in a collimated jet. 
At least some of the  pairs are likely to have a small relative velocity.  If bound
states of positronium are formed,  these atoms, which can have large inter-particle separation and long lifetimes, could travel along the jet. This would suggest that the plasma consists of  electrons, positrons, and neutral positronium atoms.  The positronium would eventually annihilate when the atom collides with hydrogen in a radio lobe or in the intergalactic medium. 

The mechanism by which a magnetic field can suppress annihilation in positronium is well understood \cite{PhysRevLett.78.199,PhysRevA.58.1129}. The physical conditions under which these states can be created are remarkably general, and one would expect there to be a myriad of astrophysical phenomena where pair production occurs in sufficiently strong magnetic fields.  
The presence of stable positronium in AGN jets is not only a feasible proposal, but is consistent with evidence that jets are pair plasmas, even though the particle density is quite high close to the AGN center.  Detection of these states would require identification of the redshifted $511\,{\rm keV}$ line.  

This paper is organized as follows.  In Section~\ref{positronium}, we present the quantum solution for positronium in a helical magnetic field and discuss the origin of the stabilization mechanism.  Concluding remarks are given in Section~\ref{conclusions}. 

\section{Long-lived positronium}
\label{positronium}

We consider the case of a positronium atom traveling near the axis of a jet from an AGN. The speed of such jets are estimated to be $1 < \gamma < 50$ \cite{Scarpa2002405}. We will consider first the non-
relativistic case ($\beta \ll 1$) and then discuss the relativistic case. There is great uncertainty in the magnitude of the magnetic field near the accretion disk. We use as an estimate
$B =10^{4}\,{\rm G}$ for the field strength just outside the accretion disk \cite{Tyulbashev}.  The magnetic field along the jet is helical and the azimuthal component is much greater than the z-component \cite{MNR:MNR8037}. The magnetic field decreases several orders of magnitude over a distance of light years. Because the field changes slowly, we treat $B$ as a constant and assume that the system will evolve adiabatically over large distances.  Atomic units (see Table~\ref{atomunits}) are used throughout unless explicitly stated.  
\begin{table}[ht]
\center
\begin{tabular}{ccc}
\hline
Charge & $e$ & $4.8029\times10^{-10}$ esu \\
Mass & $m$ & $9.1085\times10^{-28}$ g \\
Length & $a_0=\hbar/me^2$ & $5.2917\times10^{-9}$ cm \\
Velocity & $\alpha c = e^2/\hbar^2$ & $2.1877\times10^{8}$ cm/s \\
Momentum & $m\alpha c= me^2/\hbar^2$ & $1.9926\times10^{-19}$ g cm/s \\
Energy & $me^4/\hbar^2$ & $4.3590\times10^{-11}$ erg \\
Magnetic Field & $m^2e^3/\hbar^3$ & $2.350\times10^{9}$ G \\
Electric Field & $m^2e^5/\hbar^4$ & $5.142\times10^{9}$ V/cm \\
\hline
\end{tabular}
\caption{Conversion between atomic units and cgs units\label{atomunits}}
\end{table}

The total Hamiltonian for positronium in homogenous magnetic and electric fields is
\begin{eqnarray}
\nonumber
H_{\rm total} = \frac{1}{2}\left(\vec{p}_1 + \vec{A}_1\right)^2 &+& \frac{1}{2}\left(\vec{p}_2 - \vec{A}_2\right)^2\\
&+& \vec{E}\cdot\vec{r}_1 - \vec{E}\cdot\vec{r}_2 -\frac{1}{\left|\vec{r}_1-\vec{r}_2\right|}
\end{eqnarray}
where $\vec{p}_i$ and $\vec{r}_i$, $i=1,2$, are the momenta and coordinates of the electron and positron respectively; for this work we choose the symmetric gauge $\vec{A}_i= \vec{B}\times \vec{r}_i/2$.  Neither the total momentum $\vec{P} = \vec{p}_1+ \vec{p}_2$ nor the mechanical momentum $\vec{\pi} = \vec{P} + \vec{A}_1 - \vec{A}_2$  are constants of motion. (The eigenvalue of $\vec{\pi}$ is $2\vec{V}$, where $\vec{V}$ is the velocity of the center of mass.)
Only the eigenvalue, $\vec{\kappa}$, of the pseudomomentum operator,
\begin{equation}
\vec{\kappa} = \vec{P} - \frac{1}{2} \vec{B}\times\left(\vec{r}_1 - \vec{r}_2\right) = \vec{\pi} - \vec{B}\times \left(\vec{r}_1-\vec{r}_2\right),
\end{equation}
is a conserved quantity.  One can carry out a pseudoseparation of the internal motion and the center of mass motion \cite{PhysRevA.49.4415}. The effective Hamiltonian for the internal motion is
\begin{equation}
\label{effham}
H = p^2 + \frac{1}{4}\left(\vec{\kappa} + \vec{B}\times\vec{r}\right)^2 - \frac{1}{r} +  \vec{E} \cdot \vec{r},
\end{equation}
where $ \vec{r} = \vec{r}_1 - \vec{r}_2$ and $\vec{p}$ are the coordinate and momentum associated with the relative motion.  The effective potential depends parametrically on the eigenvalue $\vec{\kappa}$ . In the case
of a magnetic field only, the ionization threshold is $I = B$.

The electric field can be eliminated from the Hamiltonian by defining an effective
pseudomomentum $\vec{\kappa}^\prime = \vec{\kappa}-2\vec{v}_d$, where 
\begin{equation}
\vec{v}_d = \frac{\vec{E}\times\vec{B}}{B^2}
\end{equation}
is the classical drift velocity.
The Hamiltonian for the relative motion is identical to Eq.~\ref{effham} with $\vec{E}=0$, except that $\vec{\kappa}$ is replaced by $\vec{\kappa}^\prime$ and there is an additional constant $\vec{\kappa}^\prime\cdot\vec{v}_d+ v_d^2$; the ionization threshold is also shifted by the same amount, $I = B + \vec{\kappa}^\prime\cdot\vec{v}_d+v_d^2$, so the condition for binding is unaffected by the shift.

\begin{figure}
   \centering
   {\includegraphics[width=2in]{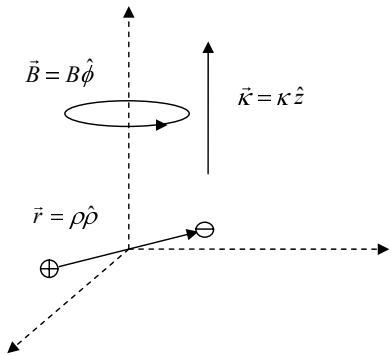} }
   \caption{A positronium atom is oriented in the $x$-$y$ plane, moving parallel to the axis of the jet.  As the system evolves in time, both $\vec{r}$ and $\vec{\pi}$ change direction, but $\vec{\kappa}$ is a constant of motion.  The magnetic field is dominated by the azimuthal component, $\vec{B} = B \hat{\phi}$.}
   \label{coordinates}
\end{figure}

We now consider the particular case where the electric field is zero and $\vec{B} = B \hat{\phi}$ . In
 general, $\vec{\kappa} = \kappa_\rho\hat{\rho} + \kappa_z \hat{z}$, but for simplicity we chose $\vec{\kappa} = \kappa \hat{z}$.  To visualize such a scenario, consider an atom which at some instant is aligned in a plane perpendicular to the $z$-axis so that $\vec{r} = \rho \hat{\rho}$; the instantaneous center-of-mass  momentum is parallel to the axis, $\vec{\pi} = \pi \hat{z}$ and $\vec{\kappa} = \left(\pi+B\rho\right)\hat{z}$, see Fig~\ref{coordinates}. The vectors $\vec{\pi}$ and $\vec{r}$ are not constrained to these instantaneous directions, but $\vec{\kappa}$ is a constant and $\kappa > 0$.

\begin{figure} 
   \centering
   \includegraphics[width=2.5in,angle=-90]{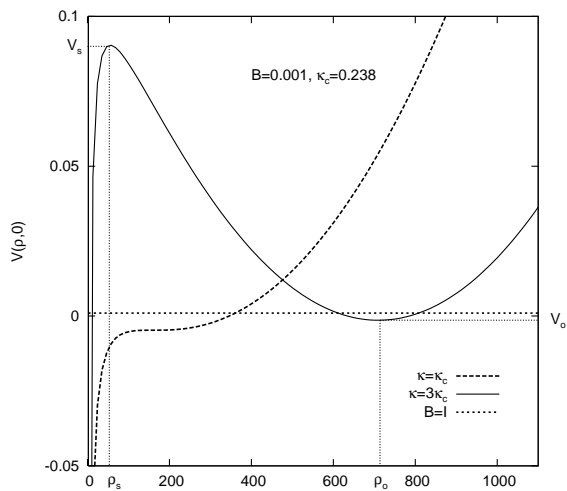} 
   \caption{The effective potential $V(\rho,0)$ is shown for $\kappa=\kappa_c$ and $\kappa = 3\kappa_c$. When $\kappa > \kappa _c$, the effective potential has a saddle point at $\rho_s$ and a minimum at $\rho_o$, in addition to the Coulomb singularity at the origin.  Bound states that are localized in the outer well can have near zero probability density at  $\rho=0$.}
   \label{figure2}
\end{figure}

In cylindrical coordinates, the Hamiltonian for the relative motion is
\begin{equation}
\label{hamiltonianregular}
H=p^2+\frac{B^2\rho^2}{4} - \frac{B\kappa \rho}{2} - \frac{1}{\sqrt{\rho^2+z^2}}+ \frac{\kappa^2}{4}
\end{equation}
The ionization threshold in the magnetic field is $I=B$, and the ionized particles escape parallel to the $z$-axis. If 
\begin{equation}
\kappa>\kappa_c = \sqrt[3]{\frac{27B}{2}},
\end{equation}
the effective potential $V(\rho, z)$ has two stationary points in addition to the Coulomb singularity.  There is a saddle point at $\rho=\rho_s$, $z=0$ and a local minimum at $\rho=\rho_o$, $z=0$ see Fig.~\ref{figure2}.  Since $\rho_o>\rho_s$, a barrier separates the outer well from the Coulomb well.  

\begin{figure} 
   \centering
   \includegraphics[width=2.5in,angle=-90]{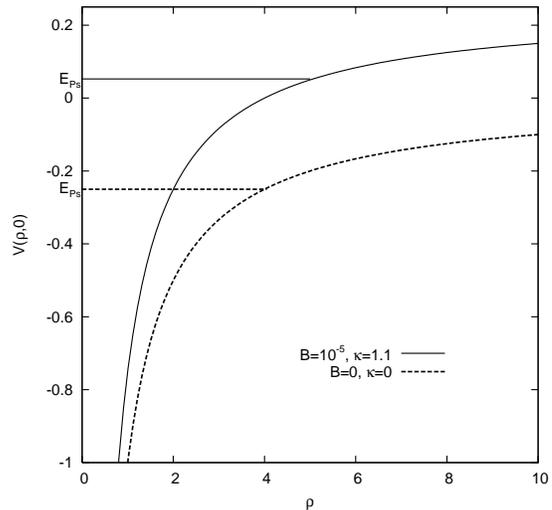} 
   \caption{In the presence of a magnetic field, the ground state energy of positronium in the Coulomb well is pushed above the ionization threshold; when this occurs, all of the bound states must reside in the outer well.}
   \label{figure3}
\end{figure}

\begin{figure}
   \centering
   \includegraphics[width=2.5in,angle=-90]{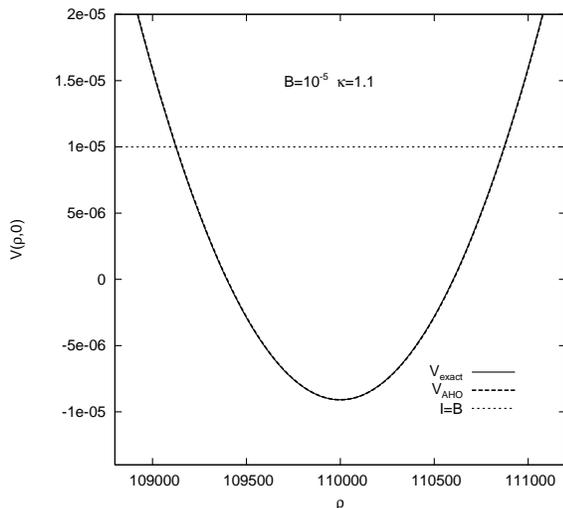} 
   \caption{The anharmonic oscillator approximation is compared to the exact potential at $z=0$. }
   \label{figure4}
\end{figure}

\begin{figure}
   \centering
   \includegraphics[width=2.5in,angle=-90]{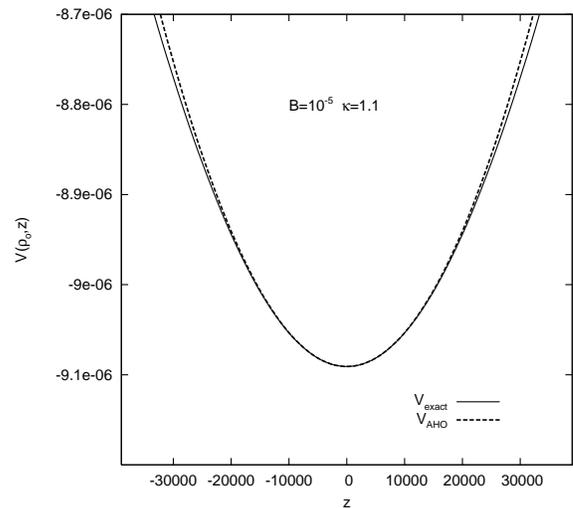} 
   \caption{The anharmonic oscillator approximation  is compared to the exact potential at $\rho=\rho_o$.}
   \label{figure5}
\end{figure}

We are in the interesting regime where $\kappa \gg \kappa_c$ (due to the large center-of-mass velocity), but the magnetic field is small, $B\approx 10^{-5}$ (atomic units).  From perturbation
theory, the ground state in the Coulomb well is
\begin{equation}
E = -\frac{1}{4} + \frac{\kappa^2}{4} - \frac{B\kappa}{2}\left<\rho\right> + \frac{B^2}{4}\left<\rho^2\right>.
\end{equation}
For $\kappa$ slightly greater than $1$, the energy lies above the ionization threshold, see Fig.~\ref{figure3}, and there are no bound states in the Coulomb well. 
(The states in the Coulomb well are subject to annihilation by the normal process.)	In order to be sure that there are no bound states in the
Coulomb well, we choose $\kappa = 1.1$. This is not unrealistic, since this corresponds to a center of mass velocity on the order of $10^{8}\,{\rm cm/s}$. In the limit where
\begin{equation}
\epsilon = \frac{B}{\kappa^2} \ll 1,
\end{equation}
the minimum and maximum occur at
\begin{equation}
\rho_o \approx \frac{\kappa}{B} = 1.1\times10^{5}
\end{equation}
and
\begin{equation}
\rho_s \approx \sqrt{\frac{2}{B\kappa}} = 430.
\end{equation}
The height of the barrier is 
\begin{equation}
V_s = V(\rho_s,0) \approx \frac{\kappa^2}{4} = 0.30
\end{equation}
 and the depth of the well is  
 \begin{equation}
 V_o = V(\rho_o,0) \approx -\frac{B}{\kappa} = - 0.91\times 10^{-5}.
 \end{equation}  The Coulomb well is separated from the outer well by an enormous barrier.  All of the bound states reside in the outer well.  At $\rho = \rho_o$, 
\begin{equation}
\frac{B^2\rho^2}{4} - \frac{B\kappa \rho}{2} + \frac{\kappa^2}{4} =0.
\end{equation}
If we expand the Coulomb potential about $z=0$, $\rho=\rho_o$, the Hamiltonian of Eq.~\ref{hamiltonianregular} reduces to that of an anharmonic oscillator,
\begin{equation}
H_{\rm AHO} = p^2 + \frac{1}{2}\frac{z^2}{\rho_o^3}+\frac{1}{2}\left(\frac{B^2}{2}-\frac{2}{\rho_o^3}\right)\left(\rho-\rho_o\right)^2-\frac{1}{\rho_o}.
\end{equation}
In Fig.~\ref{figure4} and Fig.~\ref{figure5}, we show that the exact potential is indistinguishable from the anharmonic
approximation over a very large range. 
The ground state energy is
\begin{equation}
\nonumber E = \frac{1}{2}\omega_z+\omega_p - \frac{1}{\rho_o}  \approx  B-\frac{B}{\kappa} = 0.91\times10^{-6},
 \end{equation}
which lies above $V_o$ and below $I$.  

The positronium atom is extremely large and very weakly bound.  The stabilization arises because of the height and width of the barrier. The width of the barrier at $V = 0$ is on the order of $10^5$. The probability that the particles tunnel through this enormous barrier is essentially zero. Since the decay rate for positronium is directly proportional to the probability density evaluated at contact, annihilation is suppressed and the positronium atom can travel along the jet. As the magnetic field gets weaker, the positronium atom gets larger and the binding energy decreases, though the system remains bound.


We now turn our attention to the relativistic case.  The mechanical momentum $\vec{\pi}$ is not constant, but we can still carry out an instantaneous Lorentz transformation to the frame in which the center of mass of the electron-positron is at rest.

The electromagnetic four-vector is given by $A_o = 0$ and $\vec{A} = (Bz\hat{\rho}-B\rho \hat{z})/2$.                                                                 
In the rest frame of the positronium (where $\vec{\pi} =0$), the magnetic field is $\vec{B} = \gamma B \hat{\phi}$;  in addition, there is now an electric field $\vec{E} = - \gamma \beta c B \hat{\rho}$ (Note that the electric field is in the opposite direction of the Coulomb field.) The drift velocity is 
\begin{equation}
\vec{v}_{\rm d} = \frac{\vec{E}\times \vec{B}}{B^2} = -\beta c \hat{z}
\end{equation}
and the effective pseudomomentum is
\begin{equation}
\vec{\kappa}^\prime  = \vec{\kappa} - 2\vec{v}_{\rm d} = \left(\gamma B \rho + 2 \beta c\right)\hat{z}.
\end{equation}
The Hamiltonian is shifted by a constant amount,
\begin{equation}
\Delta E  = \vec{\kappa}^\prime \cdot \vec{v}_{\rm d} + v_{\rm d}^2 = -\kappa^\prime \beta c + \beta^2c^2.
\end{equation}
The ionization energy is $I=\gamma B + \Delta E$ and the outer well minimum is
\begin{equation}
V_o \approx - \frac{\gamma B}{\kappa^\prime}+ \Delta E.
\end{equation}
The ground state energy is 
\begin{equation}
E = \gamma B - \frac{\gamma B}{\kappa^\prime} + \Delta E,
\end{equation}
which still satisfies $V_o < E < I$.  The basic mechanism for stabilizing positronium in unchanged even at relativistic speeds.

\section{Discussion}
\label{conclusions}

The stable states of positronium will undergo annihilation as they reach the end of the jet.  Some of the positronium  might undergo annihilation earlier, due to shocks in the jet \cite{AA.25.2001}; these events are likely to be undetectable since they are buried in a non-thermal background.  Nevertheless, there is overwhelming evidence that many of the jets survive long distances (hundreds to thousands of kiloparsecs) and terminate as they encounter the intergalactic medium or a radio lobe.

Stable states of positronium that enter the intergalactic medium will eventually collide with hydrogen.  In the field free case, the positronium atom is on the order of one atomic unit.  However bound states in the presence of a magnetic field have an average size of $\rho_o \approx\kappa/B$, which can be many orders of magnitude larger.  This will dramatically increase the total positronium-hydrogen scattering cross section, even at high energies.  Although the particle density is low, $n\sim10^{-3} \,{\rm cm^{-3}}$ \cite{Freeland}, the column density over just one kiloparsec is  $Q\sim10^{18}\,{\rm cm^{-2}}$.  (The particle density in the radio lobe is even greater than that of the intergalactic medium.) Any scattering process will disrupt the conditions for the stabilization of the positronium atom, leading to ionization or annihilation.   

When detecting the photons produced during the annihilation of the stable positronium atom, two effects must be considered: (1) a doppler shift due to the change in coordinate system between the observer and the emitted $511\,{\rm keV}$ photon and (2) a redshift due to the apparent recession of the AGN, $z$.  The observed frequency for an annihilation that occurs between a positron and electron, with approximately the same velocity, is
\begin{equation}
\nu_{\rm obs} = \frac{1}{1+z}\sqrt{\frac{1+\beta}{1-\beta}}\left(\frac{511\,{\rm keV}}{h}\right).
\end{equation}
Using $\gamma \approx 10$, $\beta \approx .995$ and $z=1$ we estimate a frequency of  $1.23\times 10^{21}\,\rm{Hz}$.  If the positronium atom is ionized (in the jet or as it reaches the intergalactic meduim), the positron can annihilate with free electrons  or via positronium-formation in a positron-hydrogen collision. The deceleration and annihilation processes have been thoroughly studied in relation to solar flares, and the relevant cross sections have been calculated \cite{murphyetal}. 

If electrons and positrons--as a pair-plasma--exist near the accretion disks of AGN,  they can be accelerated by rotating, helical, magnetic fields and ejected from the force-free zone.  The conditions at the Alfv\`en radius allow for the formation of stable positronium. The atoms will travel long distances along the jet and annihilate when they encounter the intergalactic medium.  The detection of red-shifted photons  would be strong evidence that these states exist.  Independent of this detection, we suggest that the most convincing argument supporting the existence of these states is that the positrons and electrons survive  the region near the Alfv\`en radius where the particle density is high.

\section{Acknowledgements}
We thank C.~M.~Urry, R.~Ross, J.~Isler, and P.~Turner for useful discussions.  JTG would like to thank the College of the Holy Cross for their hospitality while some of this work was being completed.  Research at the Perimeter Institute for Theoretical Physics is supported by the Government of Canada through Industry Canada and by the Province of Ontario through the Ministry of Research \& Innovation.  This work was partially supported by the National Science Foundation through a grant for the Institute for Theoretical Atomic, Molecular and Optical Physics (ITAMP) at Harvard University and Smithsonian Astrophysical Observatory.
JS is grateful for the hospitality and resources provided by the ITAMP Visitor Program.

\bibliographystyle{unsrt}
\bibliography{positron}

\end{document}